**Heating Mechanisms for Intermittent Loops in Active Region Cores from AIA/SDO EUV Observations**

A.C. Cadavid, J.K. Lawrence, D. J. Christian, D.B. Jess, and G. Nigro

**Abstract**
We investigate intensity variations and energy deposition in five coronal loops in active region cores. These were selected for their strong variability in the AIA/SDO 94 Å intensity channel. We isolate the hot Fe XVIII and Fe XXI components of the 94 Å and 131 Å by modeling and subtracting the "warm" contributions to the emission. HMI/SDO data allow us to focus on "inter-moss" regions in the loops. The detailed evolution of the inter-moss intensity time series reveals loops that are impulsively heated in a mode compatible with a nanoflare storm, with a spike in the hot 131 Å signals leading and the other five EUV emission channels following in progressive cooling order. A sharp increase in electron temperature tends to follow closely after the hot 131 Å signal confirming the impulsive nature of the process. A cooler process of growing emission measure follows more slowly. The Fourier power spectra of the hot 131 Å signals, when averaged over the five loops, present three scaling regimes with break frequencies near 0.1 min$^{-1}$ and 0.7 min$^{-1}$. The low frequency regime corresponds to 1/$f$ noise; the intermediate indicates a persistent scaling process and the high frequencies show white noise. Very similar results are found for the energy dissipation in a 2-D "hybrid" shell model of loop magneto-turbulence, based on reduced magnetohydrodynamics, that is compatible with nanoflare statistics. We suggest that such turbulent dissipation is the energy source for our loops.

**1.Introduction**

One approach to understanding coronal heating has been to study the thermal behavior of coronal loop structures found above and near solar active regions (ARs). Several studies indicate that "warm loops" on the periphery of AR cores, with temperatures around 1 MK, have narrow temperature distributions (Aschwanden & Nightingale 2005; Bradshaw 2008; Tripathi et al. 2009) and that they evolve in time from hotter to cooler temperatures (Ugarte-Urra, Warren & Brooks 2009; Mulu-Moore, Winebarger & Warren 2011). In contrast, structures in the AR core, including the loop apexes (Warren, Brooks & Winebarger 2011) and inter-moss areas between moss patches of opposite polarity (Warren, Brooks & Winebarger 2012), have been found to remain steady over periods of hours**.** Earlier studies of magnetic moss regions (Antiochos et al. 2003) have found steady heating, but recent observations with the High Resolution Coronal Imager (Hi-C) indicate high variability in moss regions at the footpoints of hot coronal loops (Testa et al. 2013). In other work, Ugarte-Urra, Warren & Brooks (2009) and Viall & Klimchuk (2011, 2012) have found that in the AR core the loop structures are not steady and that they evolve from hotter to cooler temperatures.

Much research has focused on relating observations to heating via impulsive bursts or "nanoflares" (Klimchuk 2006 and references therein). The bursts might be the result of reconnection events among braided fields as in the original nanoflare scenario of Parker (1988), or possibly due to the dissipation of magnetohydrodynamic (MHD) turbulence inside the loop structures (Nigro et al. 2004; Reale et al. 2005; Veltri et al. 2005; van Ballegooijen et al. 2011; Asgari-Targhi & van Ballegooijen 2012; Asgaro-Targui et al. 2013). Within the nanoflare model different outcomes follow from different heating rates of the sub-resolution loop strands. In a high-frequency process the time between heating impulses is shorter than the cooling times of the strands thus smoothing the changes over a period of time. This process can be referred to as "steady heating." In the case of low-frequency heating the time between bursts is longer than the cooling time of the structure, and the strands evolve individually to cooler temperatures. Such a process is labelled "nanoflare heating" (Cargill & Klimchuk 1997; Tripathi, Klimchuk & Mason 2011; Winebarger et al. 2011).

Within this class of processes is a scenario in which the loop consists of a small number of strands that are all heated at about the same time. The observed loop properties follow the time evolution of the strands in this "short nanoflare storm" (Klimchuk 2009). Alternatively the loop can be made of many strands all heated independently at random times. In this "long nanoflare storm" (Klimchuk 2009) the loop properties can appear steady due to averaging of the contributions over many strands. Actually, the problem is more subtle in that both the low and high frequency processes can dominate at different stages during the long term evolution of an AR (Ugarte-Urra & Warren 2012). Independent of the details, all nanoflare heating scenarios predict the presence of very hot plasma in the coronal loops. Although Reale & Orlando (2008) have shown the difficulty of detecting the hot plasma outside of flaring regions, a variety of studies using data from Hinode, RHESSI and the Solar Dynamics Observatory have applied various techniques to discover very high temperatures in loop like structures in such non-flaring active regions (e.g. Patsourakos & Klimchuk 2009; Reale, McTiernan & Testa 2009; Reale et al. 2009; Testa et al. 2011; Reale et al. 2011; Testa & Reale 2012; Petralia el al. 2014).

Ugarte-Urra and Warren (2014; henceforth UW14) have approached the problem of the energy deposition in ARs by investigating the statistical properties of heating events defined by AIA intensity signals in the 94 Å channel, in which the $\sim 10^{6.85}$ MK hot component from the Fe XVIII ions has been isolated. They find a minimum frequency of 2-3 heating events per hour. Using the "Enthalpy-based Thermal Evolution of Loops" (EBTEL) zero dimensional hydrodynamic coronal model (Klimchuck, Patsourakos & Cargill 2008), they investigate cases with heating events of different frequencies. These were assumed to be square pulses of random amplitudes obtained from a power law distribution, with a constant event duration of ~200s, and with times randomly chosen from a normal distribution. The results indicated that the actual frequency of the heating events can be higher than those "observed" since these synthetic intensities integrate over the contributions of a multiplicity of events.

In this paper we approach the problem of energy release into the corona by investigating the dynamics of five "hot" loops in AR cores that were initially selected for the high temporal variability of their intensities in the 94 Å channel. To investigate the properties of energy deposition and the hot loop evolution we use data from the AIA/SDO instrument (Lemen, et al. 2011 ) and consider the six Extreme Ultraviolet (EUV) channels: 131, 94, 335, 211, 193, and 171 Å. We also use contemporaneous data from the HMI/SDO instrument to define what we call the "inter-moss" region in the loop structures. From spatial averages along these loop inter-moss segments we obtained the intensity time series that we analyzed for this project.

We find that the non-steady loops of the kind under study are impulsively heated in a mode compatible with a nanoflare storm. Abrupt loop intensity brightenings typically begin with a sharp impulse in the hot 131 Å intensity followed in sequence by progressively broader, smoother and later increases in the hot 94 Å intensity and then the 335 Å intensity. When there is enough time between heat impulses, the cooler signals in the 211 Å, 193 Å and 171 Å are also observed to increase in that progressive order from hotter to cooler. A sharp increase in loop electron temperature tends to follow quickly after the hot 131 Å signal confirming the impulsive nature of the process. Since the hot 131 Å intensity is the least affected by processes within the loop, such as cooling and plasma flows, it is our best choice as a proxy for energy input. The power spectra of the hot 131 Å intensity fluctuations and the energy dissipation in an MHD turbulence model (Nigro, et al. 2004; Reale, et al. 2005) both show the characteristics of a $1/f$ process for the lower frequencies, "strong persistence" for the intermediate regime, and white noise at higher frequencies. This reaffirms the use of the hot 131 Å intensity to investigate the energy release process in the corona, and it suggests that the energy source for the observed intensity increases is dissipation from MHD turbulence. Based on the location of the brightenings we conjecture that the energy release is occurring at the loop tops.

In Section 2 we introduce the observations and techniques used to prepare the data for later analysis, including the identification of the hot signals. In Section 3 we describe how the loops are identified, including our background subtraction method, and the definition of the intensity time series. Section 4 presents the analysis of the patterns of loop intensity fluctuations and the comparison to modeled results. Section 5 details our loop heating scenario, and Section 6 presents the scaling properties of the loop energy input. Lastly, in Section 7 we summarize our findings and further discuss the implication of the results.

## 2. Data

We use data in the six Extreme Ultraviolet (EUV) channels (131, 94, 335, 211, 193, 171) Å from the Atmospheric Imaging Assembly (AIA) (Lemen et al. 2011) on board the Solar Dynamics Observatory (SDO) (Pesnell et al. 2012). This is in virtually continuous operation and covers the full solar disk with 0″.6 (~ 0.44 Mm) pixel scale (spatial resolution of 1″.2 or ~ 0.9 Mm) and a cadence of 12 s. The Helioseismic and Magnetic Imager (HMI) instrument on board SDO provides contemporaneous, full-disk, line-of-sight, photospheric magnetic data at a cadence of 45 sec. These locate sites of underlying magnetic moss so that their fluctuations can be excluded from the observations of loop intensity changes.

The class of events under study was discovered by means of animations of series of AIA/SDO observations of NOAA AR 11250 during its first disk passage. The observations were made on 2011 July 13, 14, 15. On July 14 AR 11250 crossed the solar meridian at S27. The Heliophysics Events Knowledgebase (http://lmsal.com/hek/hek_isolsearch.html) reports no observations of flares in this AR during the time intervals we are studying. Likewise, no GOES flares are reported during these times. The data did reveal a number of transient 94 Å brightenings along short (~ 20-30 Mm) magnetic loops in the AR core that connect areas of opposite polarity moss. Here we present the detailed analysis of the times series of length 270 min for five loops labelled A-E. Table 1 gives their dates, starting times, length of the loops, and the boundaries of the inter-moss regions. Figure 1 shows images in the 94 and 131 Å signals for Loop D at ~ 50 min after the starting time of the series at UT 12:02.

Table 1. Temporal and spatial properties for loops A-E.

| Loop | Date | Starting Time | Length (Mm) | Inter-Moss Limits (Mm) |
|---|---|---|---|---|
| A | 15-Jul-11 | UT 11:32:00 | 30.5 | 10.9 - 23.9 |
| B | 14-Jul-11 | UT 12:02:00 | 32.2 | 13.0 - 21.8 |
| C | 13-Jul-11 | UT 12:02:00 | 24.8 | 7.8 - 15.7 |
| D | 13-Jul-11 | UT 12:02:00 | 33.5 | 16.5 - 26.1 |
| E | 13-Jul-11 | UT 12:02:00 | 23.9 | 10.9 – 19.6 |

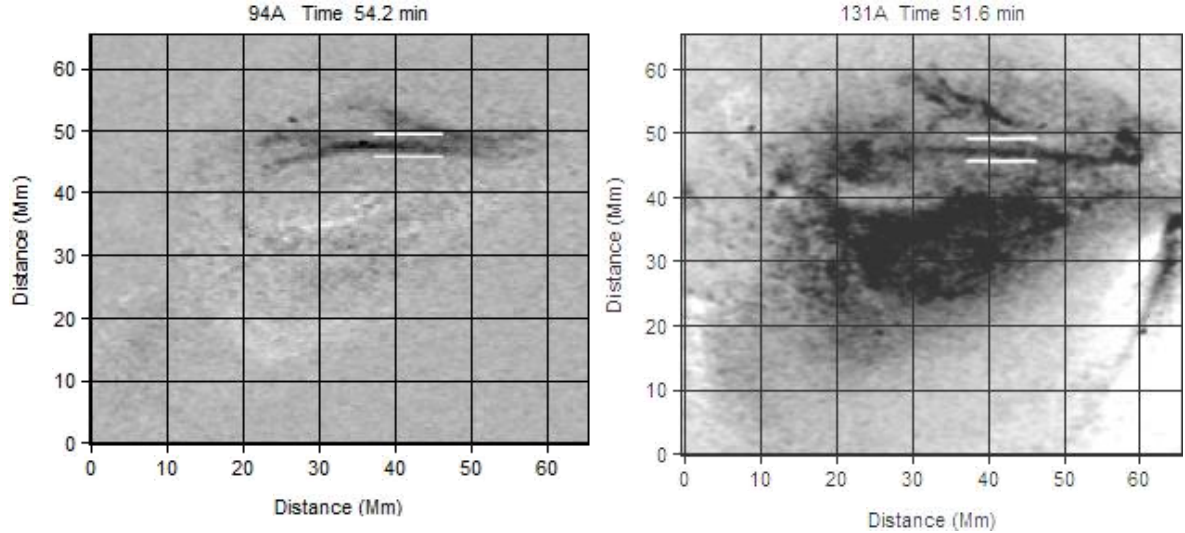

**Figure 1**. Images of AR11250 on 13 July 2011 in the 94Å and 131Å channels taken at times 54.2 min and 51.6 min respectively, measured from UT 12:02:00. The grayscale is reversed. The white lines indicate the Loop D inter-moss region studied.

To investigate energy deposition in the AR we isolate the "hot" components of this channel and also of the 131 Å channel. The AIA 94 Å response function has two maxima: a "hot" component at T ≈ $10^{6.85}$ K, due to the Fe XVIII ion, and a "warm" component at T ≈$10^{6.2}$ K (Boerner et al. 2014). Reale et al. (2011) extract the hot 94 Å signal by subtracting a fraction of the 171 Å signal that is used to model the "warm" component (which these authors label "cool"). Warren, Winebarger & Brooks (2012) and UW14 model the warm component of the 94 Å emission in terms of a combination of the 171 Å and the 193 Å emission by introducing a quartic polynomial in the quantity $x = f I_{171} + (1 - f) I_{193}$. Our variant of the approach to defining the hot component of the signal is to take it to be $I_{94h} = I_{94} - a x$. To find the factor $f$ and the coefficient

$a$ we use trial-and-error to minimize the absolute value $|I_{94h}|$ averaged over regions away from the AR core and over all times in the 270 min data block. For example, for the observations in 15 July 2011 we obtain $f = 0.19$ and $a = 0.0051$. $a = 0.0051$. We note that Del Zanna (2013) found an estimate of the Fe XVIII contribution to the 94 Å signal by modeling the "warm" component as a linear combination of the 171 and the 211 Å bands. The 131 Å response function also has two maxima: a "hot" component at $T \approx 10^{7.05}$ K. corresponding to emission by Fe XXI ions, and a "warm" component centered at $T \approx 10^{5.75}$ K. The relatively cool temperature of the "warm" part of the 131 Å response function overlaps only the 171 Å response function. In this case we model the hot emission as $I_{131h} = I_{131} - a\, I_{171}$, and we implement a similar minimization procedure to that for the 94 Å case. We obtain a value of $a = 0.044$ for the 15 July 2011 data. The values of the scaling factors have variations depending on the quiet region away from the AR core chosen to perform the analysis. For the 131 Å case one of the lowest scaling factors encountered was $a = 0.032$. Within these limits we find that the choice of factor does not significantly affect results presented later in the paper. The $a = 0.044$ choice is near the ratio of the filter response functions which gives 0.047. In the case of the 94 Å channel the ratio of the filter response functions leads to $a = 0.0071$. These models for the warm components, especially the 131 Å case, are by necessity coarse approximations. However they appear quite effective in allowing us to separate the approximate contributions of the hot Fe XVIII (93.93 Å) and the hot Fe XXI (128.75 Å) which dominate the emission in the signals during the impulsive phases of the loop brightenings. We have found no indication of a hot component in the 193 Å bandpass representing temperature $\approx 2 \times 10^7$ K. Figure 2 displays time traces of the 131 Å and 94 Å "total" and "hot" signals for the background-subtracted and spatially averaged inter-moss segment of Loop A (which will be defined in the next section and which we will use as a prototype). In the 94 Å intensity time trace the hot component dominates throughout this signal. In contrast, in the 131 Å intensity time trace the hot signal dominates the first brightening but the warm contribution is large at other times and dominates the wide intensity peak centered at ~ 56 min.

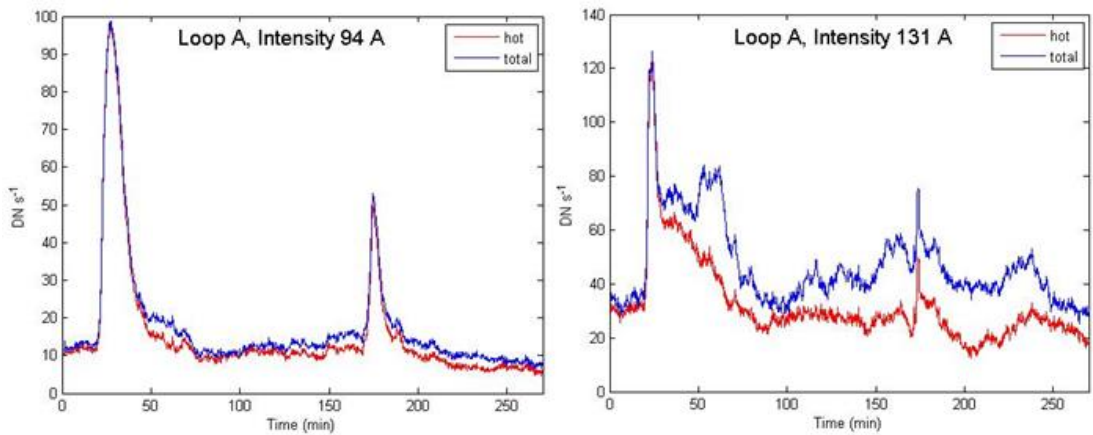

**Figure 2.** Left: the time trace of the hot 94Å intensity with background subtracted and averaged over the "inter-moss" segment of Loop A compared to the trace after subtraction of the warm portion of the 94Å response function. Right: The equivalent traces for the hot 131Å intensity. (A color version of this figure is available in the online journal)

The loops are defined by the brightenings in the 94 Å images in a non-flaring region. To estimate the energy content and compare it to the flare luminosity scale we have applied the following procedure. For a given time series of images we selected a 512 x 512 pixel region which included the AR under study and summed the intensity over all pixels. From this "space integrated" intensity time series we then subtracted a background level so that each value is the space integral of the emission in the image. We then identified the maximum value in the time series as a measure of the energy in the emission event. To calibrate the energy measure we applied the procedure to two time series which included flares. The first event was the GOES C1.2 flare on 2011 March 3, UT 19:24-19:44 which was thoroughly studied by Petkaki et al. (2012) and who found that the 94 Å emission band was dominated by the Fe XVIII line and the 131 Å band by the Fe XXI line, which correspond to the hot contributions. Therefore to avoid over processing we have used the original AIA observations. The peak integrated intensity had a value of 1,824,180 DN/s in the 94 Å channel and 3,070,000 in the 131 Å channel. For comparison we also considered the GOES B3.3 flare on 2011 July 13, UT 00:47-00:54, and the corresponding peak integrated intensities were 390,000 (94 Å) and 1,260,000 (131 Å). The procedure was then applied to AR 11250 and we found the following ratios between the peak integrated intensities of the flares and those of the brightenings for loops A and B. For the 94 Å channel: C1.2/B3.3=4.7, C1.2/Loop A= 11.6, C1.2/Loop B = 19.5. For the 131 Å channel: C1.2/B3.3=2.4, C1.2/Loop A= 21.1, C1.2/Loop B = 39.0. In summary, compared to a C1.2 flare the larger brightenings in our loops have an intensity which is down by a factor between 10 and 20 for the 94 Å cannel, and between 20 and 40 for the 131 Å channel. The brightenings are down from the B3.3 flare intensity by factors between 2.5 and 4 for the 94 Å case, and between 9 and 16 for the 131 Å case.

## 3. Loop Identification and Definition of the Inter-Moss region

We define the "spine" of a loop by first identifying the 94 Å image in the time series with maximum loop intensity and then averaging over the 11 images centered on this image in time. Then for each pixel coordinate in the EW-direction we find the pixel NS-coordinate with maximum average intensity. These EW- and NS- coordinates of the spines identify loops that are projected, both in the line-of-sight and in the NS-directions, through time. For each image and at each EW-pixel of the spine we average over ± 4 pixels (total width of 3.9 Mm) from the spine in the NS-direction. Thus, at each time we are averaging over an area on the Sun that will (ideally) enclose the whole loop feature, or at least the loop core, through time in the segment of interest. It can accommodate NS loop motion up to 2 Mm in either direction. The averages also will contain significant non-varying background. The same spine coordinates are used to calculate the loop intensities in the remaining five EUV time series leading to six space-time maps for each loop. Figure 3 maps the loop spine intensities showing the evolution of the five loops in the 94 Å channel. The "brightenings" characterized by the rapid increase in intensity will be thoroughly studied in the following sections.

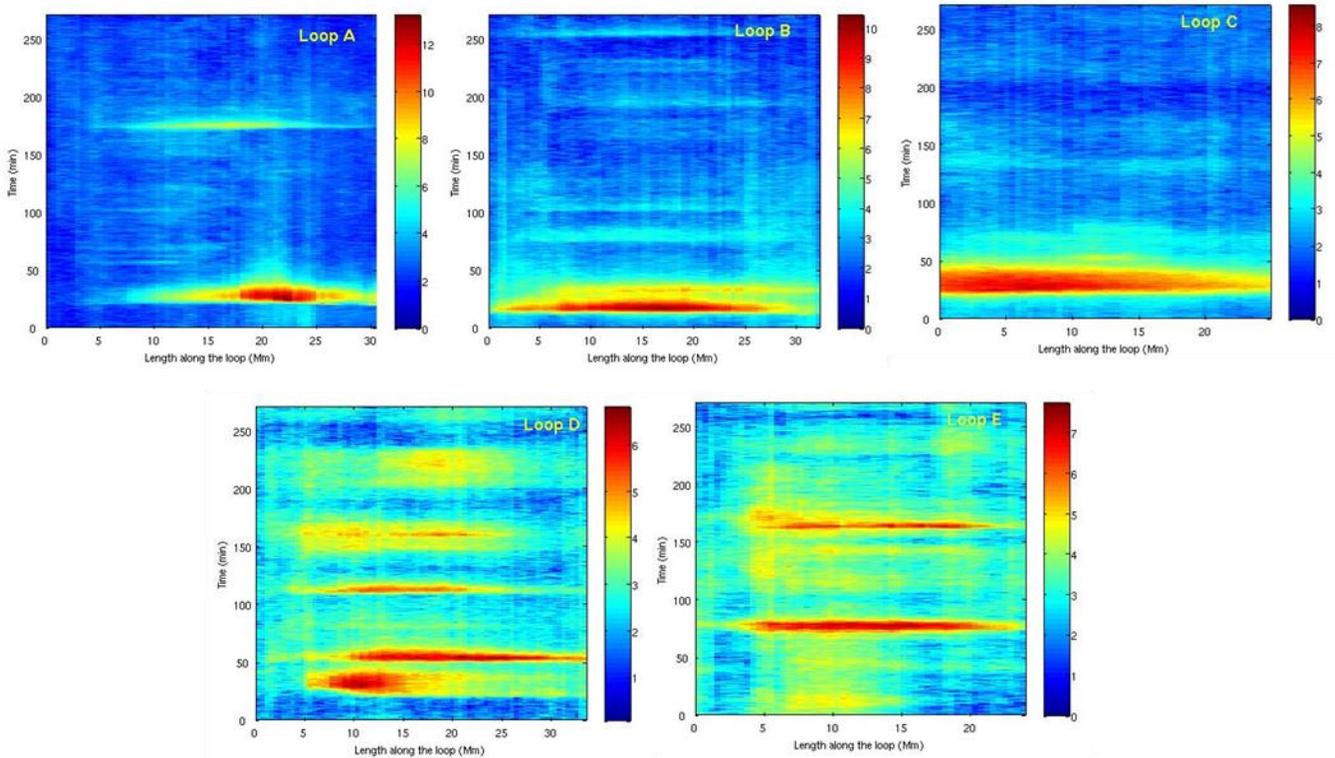

**Figure 3.** Space-time maps of the background-subtracted 94 Å intensity for loops A-E. To better display the full range of values the square root of the intensity is presented. (Intensity units are DN s$^{-1}$). (A color version of this figure is available in the online journal)

In order to reveal the loop intensity changes we must separate the variable loop signals from the steady background emission. It is well-known that the method used for background subtraction can lead to a noticeable effect on results (e.g. Del Zanna & Mason 2003; Terzo & Reale 2010; Aschwanden & Boerner 2011). An established technique appropriate for steady loops was developed in Klimchuk 2000 in which the background is identified by interpolating across the axis of the loop. In another approach Aschwanden & Boerner (2011) and Aschwanden et al. (2013) perform a fit across each loop point which includes a Gaussian profile and a cospatial linear background. These approaches are based on attempting a spatial isolation of the loop emission from that of underlying plasma.

Our objective here is to investigate the temporal behavior of a particular class of loops. The hot intensities display a variably emitting component superimposed upon a background of steady emission. Thus, we need to concentrate on the variable component and remove the effect of the constant background emission, which may partly reside within the loop itself. So we try a separation based on variability in time rather than in space. For each pixel position along the loop spine, we find the minimum intensity value there through time, and then subtract that minimum value from all values at that pixel location. This procedure is designed to avoid any negative background-subtracted intensity values. But, partly because of noise, the minimum values along a given loop spine come at different times. Thus the intensity averaged over a finite loop segment will show a significantly positive minimum value. See Figure **5** below.

The subtraction procedure is applied to the data in the six EUV channels. In the study of the dynamics of a small flare Petkaki et al. (2012) perform a similar temporal background subtraction selecting the value of the lowest total emission in the 94 Å and 131 Å channels. Figure 3 maps the background-subtracted 94 Å intensity along the five loops through time. Loops A and B are characterized by one main "concentrated" brightening with very small subsequent intensity peaks. Loop C exhibits a brightening "interval" with more structure. Loops D and E show multiple brightenings of varying intensity.

Inspection of Figure 3 shows that the major increases in the 94 Å intensity occur around the apex of the loop. In order to identify this segment in the loops we have calculated space time maps of the contemporaneous photospheric line-of-sight (LOS) magnetic field observed with the HMI instrument. The strong positive and negative LOS fields corresponding to the moss regions in the photosphere are localized at distances < 5 Mm and > 25 Mm along the projected Loop A as seen in Figure 3 (top left). To further refine the limits we have averaged the magnetic field in time. We then define the inter-moss region in the photosphere by the boundaries where the *average* value of the LOS photospheric field is small. This corresponds to ~11 Mm and ~24 Mm along Loop A as seen in Figure 3. There are variations among loops. For example for loops B and D the average photospheric LOS magnetic field is essentially negligible in the photospheric inter-moss region, while for loop E there is no plateau between the two polarities so the boundaries of the inter-moss region are estimated purely from the space-time intensity maps. Observations that are superimposed on moss show different properties than those we are presenting in this paper. Although they do not reflect the chromospheric or coronal fields, we use the boundaries identified for the photospheric magnetic field to define what we will refer to as the "inter-moss" region in the loops. This is an approximation since we are dealing with the loop projection but it suffices to determine a loop segment away from the moss and near the apex region.

**4. Loop Intensity Fluctuations**

In order to follow the time evolution of the loops with individual light curves, we have averaged the intensities along the inter-moss segments listed in Table 1. In Figure 4 (left) we summarize and illustrate the effects of the process of selecting the hot component and then subtracting the steady emission background. All three traces are averaged over the inter-moss region for the Loop A, 131 Å signal. In addition to the effects already noted in Figure 2, the intensity for the hot component is about one half of that of the total signal. The background subtraction reduces the intensity level by an additional factor of four. Figure 4 (right) presents the standard error resulting from averaging over the inter-moss region for the three light curves. For the peak of the first brightening the error has essentially the same value in the three cases. For the hot background subtracted case it corresponds to about 10% of the signal. For the second peak near 170 min the error is about 5% and for the steady segments it can be as large as 8%. Figure 5 presents the resulting time series for the hot components of the 131Å and 94Å intensity light curves for the five loops. In all cases the major brightenings are characterized by a peak in the hot 131 Å signal closely followed by a peak in the 94 Å series. The time differences between the two maxima range between 0.8 min and 3.2 min with an average 1.5 min. The intensities are in the range 40-100 DN s$^{-1}$ px$^{-1}$.

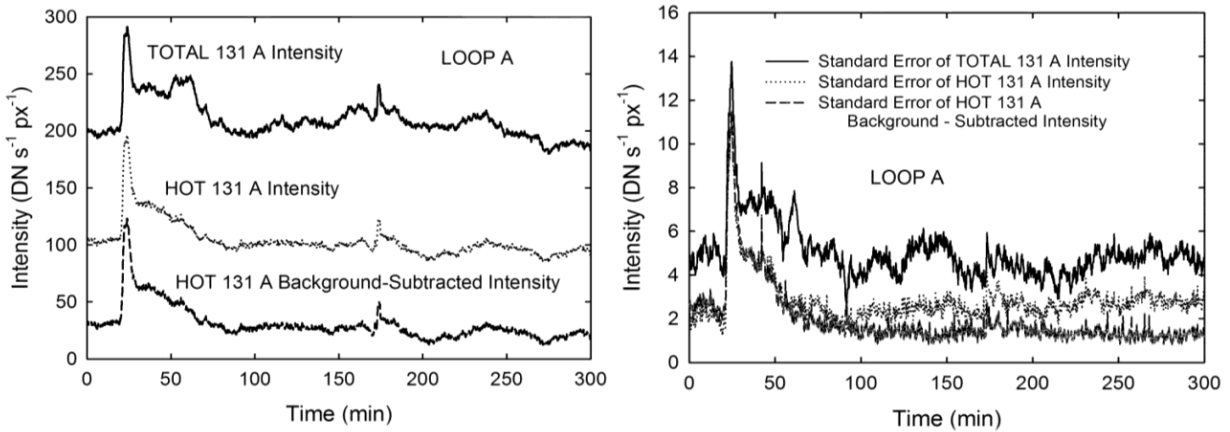

Figure 4. Left: light curves for the total, hot, and hot background subtracted 131 Å signal for Loop A after averaging over the inter-moss region. Right: the corresponding standard errors.

To investigate details of the heating mechanism we have followed the evolution of the intensity averages in all of the EUV channels. Figure 6 presents three examples of brightenings in which the energy initially released in the hot 131 Å is followed by the hot 94 Å signal and the cooler channels in progressive order (hot 131, hot 94, 335, 211, 193, 171). For the first brightening in Loop A the peak in the hot 131 signal occurs at 23.8 min, and the rest at time delays of (3.2, 14.0, 36.8, 37.0, 37.8) min. In the case of the first brightening in Loop B the hot 131 peak occurs at 15.6 min and the rest at time delays of (1.6, 8.6, 15.6, 16.8, 18.2) min, and for the second brightening in Loop C the hot 131 peak occurs at 52.4 min and the rest at time delays of (1.4, 3.4, 7.4, 7.4, 8.4) min. While there is no common value for the time difference between the peak times in the three examples, in all cases the 131 Å and 94 Å peaks are close in time. Also the 211 Å, 193 Å and 171Å are grouped together. The 335 Å signal has a wide peak and is more separated in time from the hot channels and the warm/cool channels.

The nanoflare storm model assumes that the strands comprising a loop are impulsively heated at different, independent times and with independent cooling times (Cargill 1994; Mulu-Moore et al. 2011). The net effect leads to a thermal distribution for the loop. Viall & Klimchuck (2011) have used the EBTEL (Klimchuk et al. 2008) model to simulate nanoflare storms of varying duration and intensity. One of their cases, consisting of high energy individual nanoflares with a triangular shape of 500 s duration, leads to intensity time series with the same salient characteristics as those obtained in the present examples (Figure 3b in Viall & Klimchuck (2011)). When another nanoflare storm occurs and there has not been enough time for the energy to dissipate through the cooler channels, the brightenings present partial "cascades" only in the hotter three channels. In Figure 6 the second brightening in Loop A, the second brightening in

Loop D and the first brightening in Loop E present peaks for the 131, 94, 335 Å signals at (174,174.8,182) min, (51.6,54.2,57.6) min and (77,77.6,82.8) min, respectively.

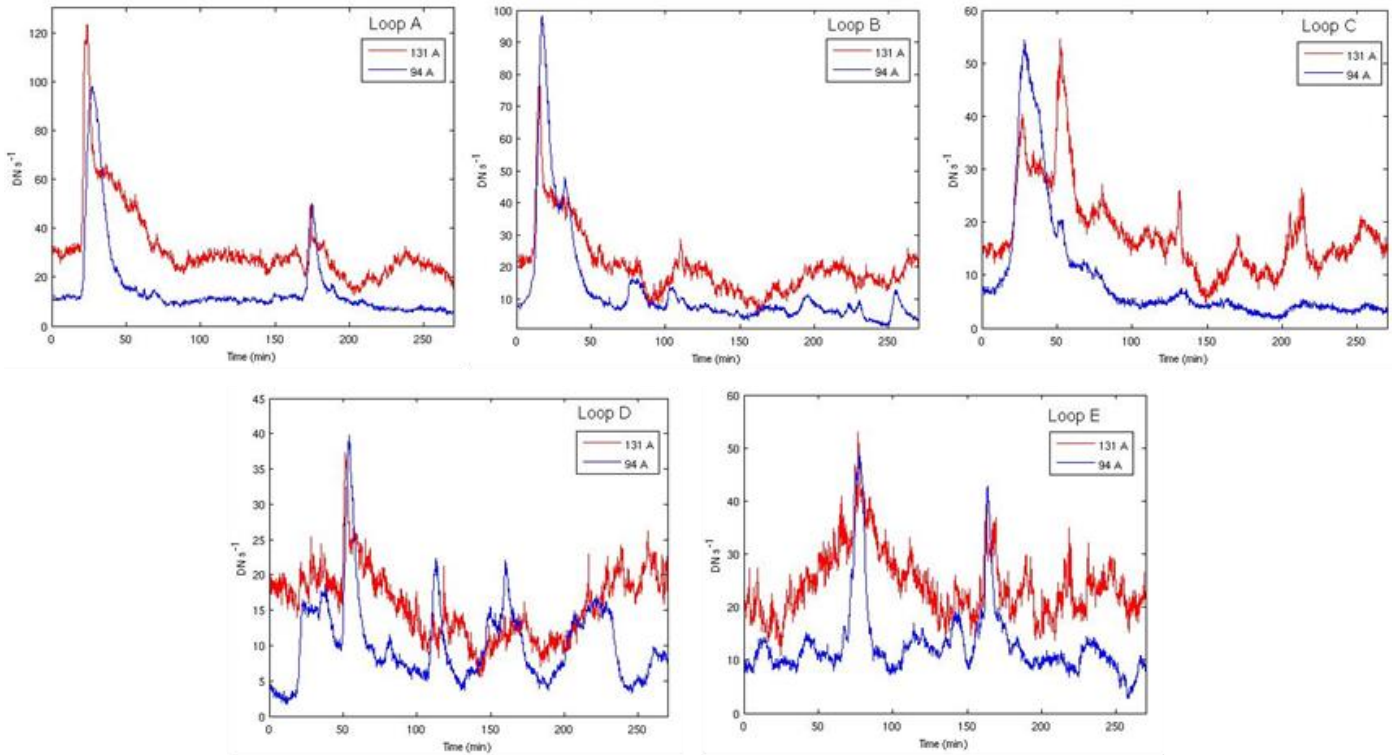

**Figure 5.** Intensity time lines of the hot 131 Å and 94 Å signals averaged over the inter-moss parts of each of the five loops. The peaks in the 94 Å channel correspond to the brightenings in the space-time maps in Figure 3. (A color version of this figure is available in the online journal)

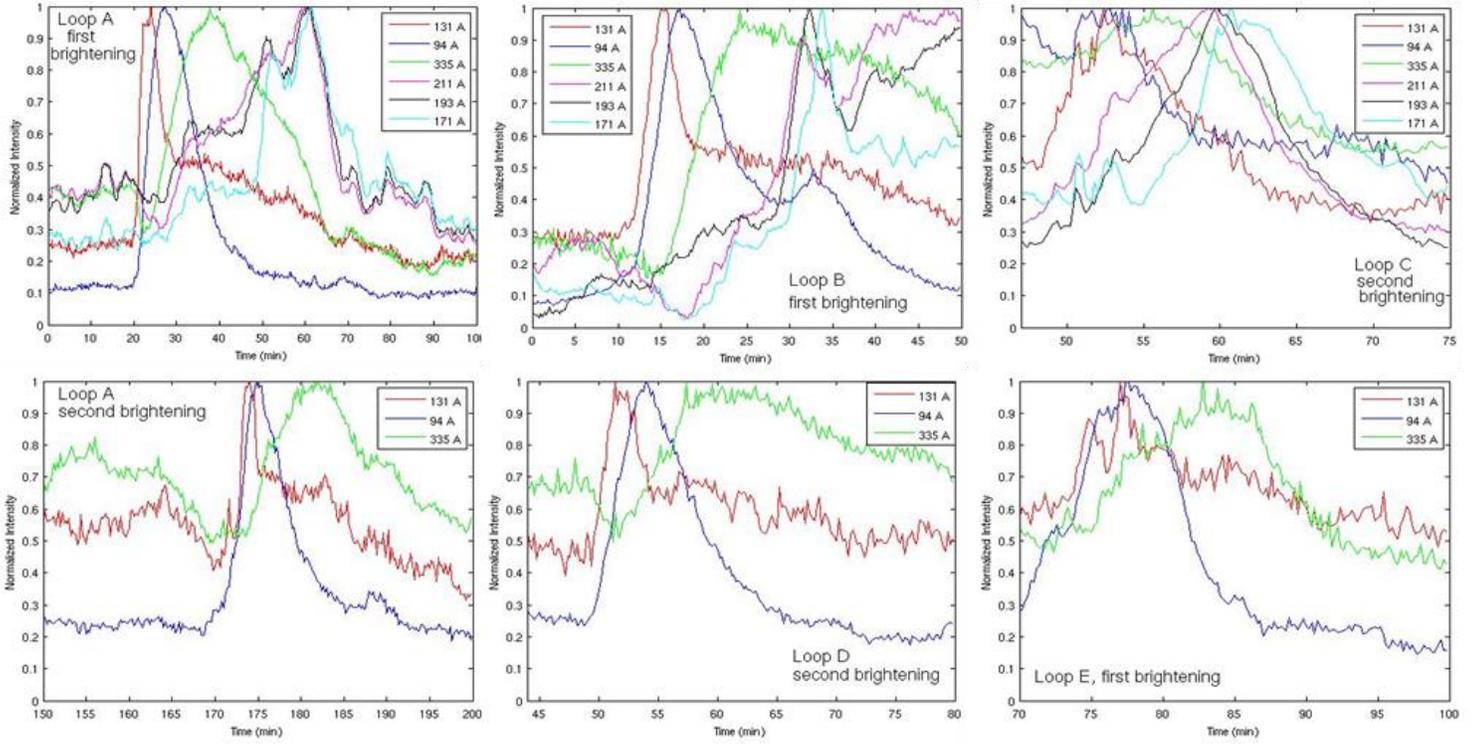

**Figure 6.** Intensity time traces around particular brightenings. Top-row: Possible nanoflare storms with progressive cooling in the six EUV channels. Bottom-row: Partial "cascade" with progressive cooling in the three hot channels. (A color version of this figure is available in the online journal)

## 5. Loop Heating Process

The results in the previous section indicate the presence of an impulsive heating mechanism. The presence of very hot plasmas is a characteristic of nanoflare heating. To gain more insight into the process here we calculate the time evolution of the electron temperature and emission measure for two characteristic heating events.

For this purpose we have adapted the method developed by Ashwanden & Boerner (2011) and Ashwanden et al. (2013). The flux in the six coronal wavelength filters can be represented by

$$F_\lambda(t,x) = \int \frac{dEM(T,t,x)}{dT} R_\lambda(T)\, dT = \sum_k DEM(T_k,t,x)\, R_\lambda(T_k)\, \Delta T_k \quad (1)$$

where the $F_\lambda(t,x)$ are the background-subtracted intensities along the loop coordinate $x$ at time $t$, and the $R_\lambda(T)$ are the instrumental response functions. We approximated the differential emission measure $\frac{dEM(T,t,x)}{dT}$ (DEM) via a superposition of Gaussian functions

$$\frac{dEM(T,t,x)}{dT} = \sum_{p=1}^{n} EM_p(t,x) \exp\left(-\frac{[\log(T)-\log(T_p(t,x))]^2}{2\,\sigma_p^2(t,x)}\right) \qquad (2)$$

A calculation at each space time point of the peak emission measures $EM_p(t,x)$, temperatures $T_p(t,x)$ and widths $\sigma_p(t,x)$, was done via chi-square optimization of a single Gaussian model and a double Gaussian model with equal amplitudes. The latter is expected to provide a more accurate analysis of the data (Reale et al. 2009; Sylwester, Sylwester & Phillips 2010), and it is particularly relevant in our case since the AR core loops, which are bright in the 94 Å, have the potential of being multi-thermal (Aschwanden et al. 2013). Because there are only six data points for each fit we have found it necessary to limit the fit to five parameters and we therefore specify equal amplitudes. We used the latest version, V4 (March 2014), of the AIA response functions obtained from SSWIDL. In the single Gaussian model the optimization was performed via a look-up table leading to an estimate of the peak quantities. For the double Gaussian model we used the Matlab routine FitChiSquare, which implements a generalized non-linear optimization according to the algorithm developed by Press et al. (1986). In contrast to the calculations in the previous references in which a fit is performed over the width of the loop and over the six intensity measurements, here we only have available the six intensities to perform the fit. Testa el al. (2012) applied the Monte Carlo Markov chain forwarding method to investigate the temperature diagnostics of synthetic AIA data and found inaccuracies due to the small number of constraints provided by the AIA data and the broad response functions. Therefore the quantitative results have to be taken with caution but, as will be shown, the qualitative joint behavior of the temperature and electron density provides useful information on the properties of the heating process.

We define the emission-weighted temperature (Chitta et al. 2013) as:

$$Te(t,x) = \frac{\sum_k DEM(T_k,t,x)\,T_k\,\Delta T_k}{\sum_k DEM(T_k,t,x)\,\Delta T_k} \qquad (3)$$

where $DEM$ is the differential emission measure introduced in equation 2, $T_k$ is the temperature, and $\Delta T_k = 0.05$ actually refers to intervals in the value of $\log(T)$. The sum in $\log(T)$ ranges between 5.35 and 7.35. The electron density is estimated by calculating $ne=(EM/w)^{1/2}$ where $EM$ is the emission measure obtained by integrating $DEM(T)$ over the temperature $T$ and $w$ is the loop width, which has been taken to be constant at 9 pixels or 3.9 Mm (Ashwanden et al. 2013).

To aid physical interpretation of the evolution of the loop temperatures and electron densities we focus on loops A and B. These each present an early main brightening with a narrow localization in time and so permit an uncluttered interpretation of the aftermath. Given the limited number of data points the $\chi^2$ goodness of fit values for the double Gaussian model are larger than for the single Gaussian model but there is generally good agreement between the results obtained with the two models. For Loop A using the single Gaussian approach 86% of the pixels had $\chi^2 < 4$, whereas for the double Gaussian model only 53% of the pixels had $\chi^2 < 4$. For Loop B the results were: single Gaussian 85% and double Gaussian 47%. Upon closer inspection, however, the double Gaussian method better represents physical details that are not captured by the single Gaussian model, such as the temperature increase associated with the small second brightening in Loop A (Figure **5**). Thus henceforth we will take the double Gaussian approach.

The emission-weighted temperature *Te* of Loop A obtained with the double Gaussian model reached a maximum temperature of 6.5 MK, and the maximum density occurs at the "peak time" (Aschwanden & Shimizu 2013) with maximum value of $n_p = 9.5 \times 10^9 \ cm^{-3}$ and a corresponding peak temperature $T_p = 3.8$ MK. A similar analysis for Loop B gives a maximum temperature of 11 MK, and the maximum electron density $n_p = 8.3 \times 10^9 \ cm^{-3}$ and corresponding peak temperature $T_p = 6.2$ MK. We note that since a 100% filling factor has been assumed across the loop and an offset has been subtracted the maximum density values are lower limits.

In Figure 7 (top-right) we plot the *Te-ne* phase diagram for the first brightening (0 < t < 90 min) of Loop A. We superimpose the line corresponding to the Rosner-Tucker-Vaiana (1978) (scaling law (RTV) $Te \propto n_e^{1/2}$ with the constant of proportionality fixed by assuming that the RTV relation is satisfied when *ne* and *Te* have their peak values $(n_p, T_p)$. Also indicated in the figure are the points corresponding to the peak time as well as the beginning and end times of the brightening. The system phase proceeds in time in a clockwise sense. Concentrating on the times around the brightening (9 min < t < 60 min) the *Te-ne* phase diagram shows the tendency for the data points to be above the RTV line during the heating phase and below for the cooling phase. This is precisely the behavior encountered in a hydrodynamic simulation for an impulsively heated single loop (Figure 4, Aschwanden & Shimizu 2013). Figure **7** (bottom right) shows the same behavior for Loop B. In earlier work Jakimiec et al. (1992) had investigated the density-temperature relation in a one- dimensional hydrodynamic model with a sudden energy release, which corresponds to the schematic description in Reale (2007, 2010). We differ from these analyses in that the present results have a less steep rise in temperature.

For steady loops it would be possible to apply the RTV scaling laws to obtain the heating rate as a function of the loop length and maximum temperature. Since the loops under study are not steady we must look for a proxy for the energy input by comparing the time evolution of the intensity signals with those of the emission-weighted temperature and the electron density. For this purpose we consider these quantities normalized at their maximum over the 270 min observation runs. Figure **7** (top-left) compares the normalized *Te* time series and the normalized 131 Å intensity for the first brightening in Loop A. Figure **7** (top-middle) compares the corresponding normalized *ne* time series, and the normalized 335 Å and 211 Å intensities. The 131 Å signal has a peak at 23.8 min that is followed by the temperature peak 1.8 min later at 25.6 min and the maximum density 9.6 min later at 33.4 min. For Loop B Figure **7** (bottom-row) the progression of maxima is 131 Å: *Te: ne* at 15.6:17.6: 27.6 min. While the temperature drops to its initial value during the duration of the brightening, the density increases and is controlled by the presence of the cooler plasma filling the loop as indicated by the temporal evolution of the 335 Å and 211 Å signals. Qualitatively, it is clear that impulsive heating is taking place with a fast increase and decrease in the temperature followed by a more gradual process in the density. The temperature peaks closely follow the hot 131 Å intensity maxima. This delay probably arises from inaccuracies in the inversion method, but the result nevertheless supports the notion that the hot 131Å signal is directly linked to the energy deposition and can serve as a proxy measure for the same.

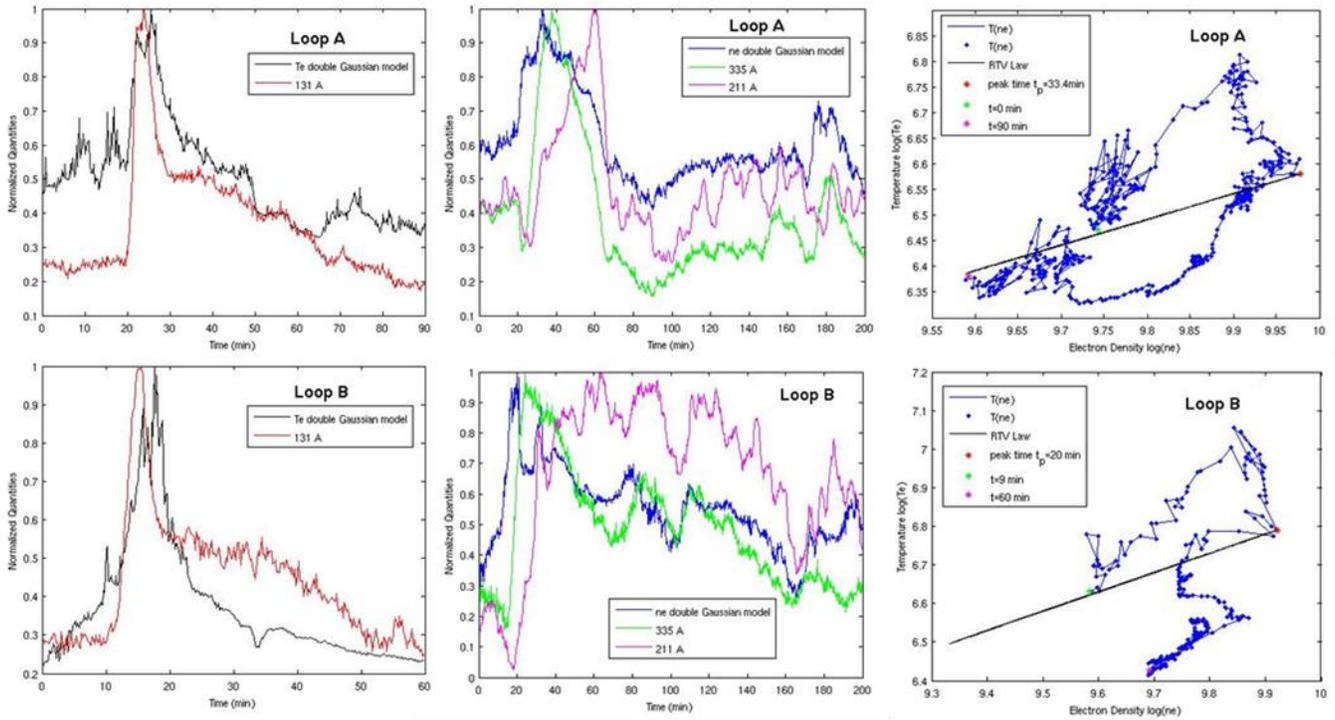

**Figure 7.** Results for Loop A are presented in the top row and for Loop B in the bottom row. Left: comparison of the time evolution of the normalized emission weighted temperature and the hot 131 Å intensity. Middle: comparison of the time evolution of the normalized electron density, and the 335 Å and 211 Å intensities. Right: phase diagram *Te-ne* for the duration of the brightening, 90 min for Loop A and 60 min for Loop B. The straight line describes the RTV scaling law. The points corresponding to the peak time for the maximum density as well as the beginning and end times are given for reference. . (A color version of this figure is available in the online journal)

At this point, we can further estimate how the energy released in the loops compares to that of the flares classified in Aschwanden and Shimizu (2013). We have applied their formula for the thermal energy:

$$E_{th}(t) = 3\, ne(t)\, k_B\, Te(t)\, V(t) \qquad (4)$$

where $k_B$ is the Boltzmann constant and $V$ is the flare volume. We have approximated the loop volumes by a cylinder over the inter-moss region. This leads to a peak thermal energy of $\sim 10^{27.5}$ ergs. A similar calculation for the C1.2 class flare studied by Petkaki et al. (2012) and using their published values gives a thermal energy of $\sim 10^{28.5}$ ergs. This order of magnitude difference between the thermal energies associated with the loop brightenings versus the C1.2 class flare is compatible with the ratios in peak intensities described in section 2. The thermal energy of the C class flare is about one to two orders of magnitude smaller than the values for the M class flares encountered by Aschwanden and Shimizu.

## 6. Power Law Spectra of Intensity Fluctuations

In the preceding analyses we have encountered strong evidence that the class of non-steady loops in our study are impulsively heated in a manner compatible with nanoflare storms. In a classic paper Hudson (1991) determined that the energy distribution of nanoflare type events also obeys a power law as previously encountered for flares. In general these phenomena are characterized by probability density functions with power law form (Aschwanden 2004 presents a summary of scaling exponents). Since we are dealing with non-steady processes we cannot directly relate observed intensity changes to inputs of dissipated energy (Martens, 2010). However, we find evidence that the hot 131 Å intensity signal can be useful as a proxy for the energy input. For one thing, abrupt loop intensity brightenings commonly begin with a sharp impulse in the hot 131 Å intensity followed in sequence by progressively broader, smoother and later increases in the hot 94 Å intensity and then the 335 Å intensity. The hot 131 Å intensity is the one least affected by processes within the loop, like cooling or plasma flows, and it is the best available indicator for energy input from the outside. To perform a statistically significant study of the probability density function of intensity event strengths we would need a much larger sample of loops. Acquisition of a significantly greater data base is a goal for future research. For now we consider power spectra of the various intensity time series. To compute the average power spectrum we first calculate the spectrum of background-subtracted intensity time series in a given wavelength band for each of the five loops, and then average the five spectra at each frequency. The results are then presented in log-log plots. Figure 8 (upper-left) shows the average power spectrum for the hot 131 Å intensity as well as for the total 131 Å intensity signals. The spectrum for the hot 131 signal has a $1/f^\beta$ form with two regimes. These show a scaling exponent $\beta = 1.31 \pm 0.07$ for $f < 0.1$ / min = $1.67 \times 10^{-3}$ Hz, and $\beta = 2.71 \pm 0.16$ for $0.1 < f < 0.7$ / min or $1.67 \times 10^{-3} < f < 1.17 \times 10^{-2}$ Hz. For higher frequencies we find white noise. In comparison, for $f < 0.1$ /min, the spectrum for the total 131 signal has a scaling exponent of $\beta = 1.81 \pm 0.12$. For this low frequency regime the difference between the slopes of $0.50 \pm 0.14$, corresponding to 3.6 σ, is significant. These results indicate that while the "hot" signal has the properties of "1/f noise" characterized by a scaling exponent $0.5 < \beta < 1.5$, the "total" signal includes contributions from a different process.

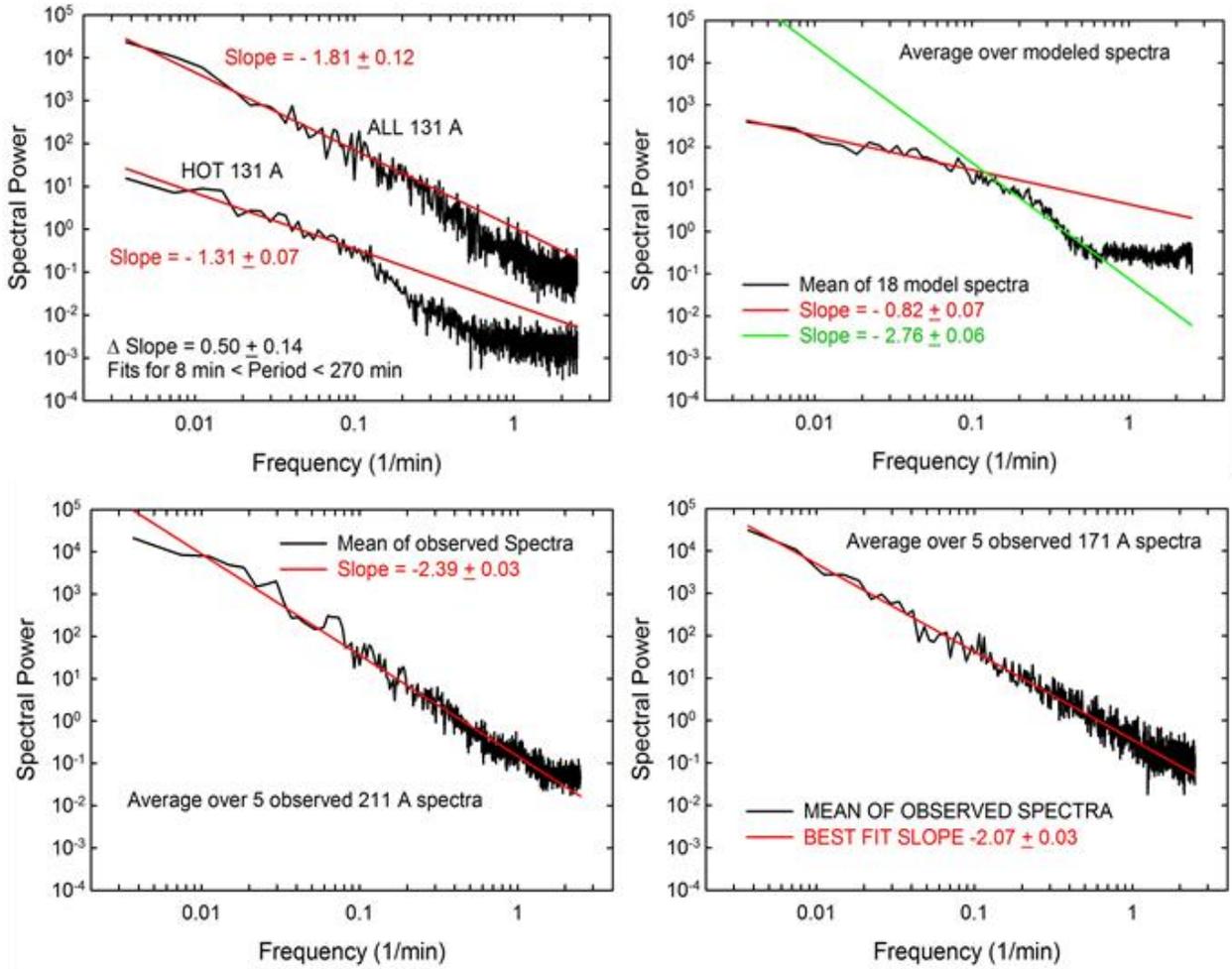

**Figure 8.** Average power spectra for different time series. Top-left: hot 131 and total 131 intensities for Loops A-E. Top-right: energy dissipation spectra from MHD turbulence model. Bottom-left: Spectra of 211 Å intensities for Loops A-E. Bottom-right: 171 Å intensities for Loops A-E. (A color version of this figure is available in the online journal)

As is well known, $1/f$ noise is encountered in many examples in nature. In particular Ueno et al. (1997) interpreted the power spectra for solar X-ray fluctuations measured by the GOES 6 satellite in terms of a superposition of flare like impulses with exponential relaxation functions $\sim \exp(-t/\tau)$ and a wide distribution of decay times $\sim (10 < \tau < 140 \text{ min})$. If the impulse response function obeys a power law $\sim t^{-\gamma}$ then the scaling exponent is given by $\beta = 2(1-\gamma)$ (Lowen & Teich 1990). In the present case of our hot 131 Å spectra this would imply a power law index $\gamma = 0.36$ applied to the range $f < 0.1/\min = 6/\text{hr}$. This includes the minimum frequency of 2-3 heating events per hour encountered by UW14. Moreover, the upper limit of the scaling range is compatible with their maximum count of $\sim 5$ events per hour (Figure 6–left in their paper). Although their nanoflare storm simulation used square impulses it is interesting to consider the possibility of a variation with pulses with power law decays. As reported in their paper, UW14 have run simulations with ~ 9 events per hour obtaining an intensity envelope which is compatible with the observations. This has led them to conjecture that the true event frequencies could be much higher. In the present case we find that the power law spectrum has a break at $f \approx 0.1/\text{min}$, and the scaling exponent changes to $\beta = 2.71$ with Hurst exponent $H = (\beta-1)/2 = 0.86$, implying the higher frequency time series has a long-term positive autocorrelation. (Schroeder, 1991).

To interpret the power spectrum with three scaling regimes we have compared our spectra to the spectra of intermittent energy dissipation derived from a numerical model in which the magnetic footpoint motions in the photosphere lead to the injection of magnetohydrodynamic (MHD) turbulence into coronal loops, where the latter is stored and dissipated (Nigro et al. 2004; Reale et al. 2005). This "hybrid" shell model, based on reduced MHD (RMHD), was used to numerically model the nonlinear couplings between magnetic and velocity fluctuations that transfer energy from large to small scales where dissipation then converts the turbulent energy to heat. The coronal nanoflares can be explained as the intermittent energy releases due to dissipation in the MHD turbulence process. The simulation does not describe the physical mechanism of magnetic reconnection, because it uses a shell model, which does not allow one to keep the details of the magnetic topology in the directions perpendicular to the main magnetic field $\mathbf{B}_0$. However it provides the capability to reach very large Reynolds numbers. In particular, the model is derived starting from the RMHD equations and performing the Fourier decomposition of the field components perpendicular to $\mathbf{B}_0$, while the dependence on the space variable $x$ along $\mathbf{B}_0$ is kept. This $\mathbf{k}_\perp$-space includes only concentric shells of wave-vectors with exponentially growing radius $k_n = k_0 2^n$ ($k_0 = 2\pi/L_\perp$, where $L_\perp$ is the width of the loop cross section). In each shell, two scalar complex amplitudes, $u_n(x,t)$ and $b_n(x,t)$, are considered for the perpendicular velocity and magnetic field, respectively. It is imposed that the nonlinear coupling of the fields amplitudes among shells conserve quadratic invariants: total energy, cross helicity, and squared magnetic potential. The evolution equations for dynamical variables $u_n(x,t)$ and $b_n(x,t)$ are hence derived in terms of the Alfvén speed (see Nigro et al. 2004 for details):

$$\frac{\partial u_n(x,t)}{\partial t} - \frac{\partial b_n(x,t)}{\partial x} = -\nu k_n^2 u_n(x,t) +$$
$$ik_n \left( u_{n+1}u_{u+2} - b_{n+1}b_{n+2} - \frac{5}{8}(u_{n-1}u_{n+1} - b_{n-1}b_{n+1}) + \frac{1}{16}(u_{n-2}u_{n-1} - b_{n-2}b_{n-1}) \right)^* \quad (5)$$

$$\frac{\partial b_n(x,t)}{\partial t} - \frac{\partial u_n(x,t)}{\partial x} = -\mu k_n^2 b_n(x,t) +$$
$$ik_n \left( \frac{1}{12}(u_{n+1}b_{n+2} - b_{n+1}u_{n+2}) + \frac{1}{6}(u_{n-1}b_{n+1} - b_{n-1}u_{n+1}) + \frac{1}{3}(u_{n-2}b_{n-1} - b_{n-2}u_{n-1}) \right)^* \quad (6)$$

with $n = 0, 1, ..., n_{max}$ ($n_{max} = 11$). Here $\nu$ is the dimensionless viscosity, and $\mu$ is the dimensionless magnetic diffusivity. The forcing for the velocity is applied to the first 3 shells ($n = 0, 1, 2$) in order to reproduce the photospheric motions concentrated at larger spatial scales $\sim L_\perp$. The forcing is realized by random Gaussian-distributed signals with an amplitude of $\sim 1$ km/s and a correlation time of $\sim 5$ minutes. This is typical of photospheric motions. Total reflection of the field amplitudes is imposed at both boundaries, that is at the two footpoints. Therefore equations 5 and 6 describe the evolution of velocity and magnetic field fluctuations $u_n$ and $b_n$ propagating along the loop, i.e. along $x$, at the Alfvén speed. On the RHS of equations 5 and 6 the nonlinear coupling terms describe the turbulent cascade of the energy that is injected at larger scales, towards the smaller dissipative scales. The model equations were numerically solved using values for the dimensionless parameters corresponding to a typical coronal loop. Therefore our numerical results reproduce the evolution of the plasma in a typical loop characterized by a longitudinal length $L = 3 \times 10^4$ km, an aspect ratio $R = L/L_\perp = 30/2\pi$, Alfvén velocity $c_A = 2 \times 10^3$ km/s, and mass density $\rho = 1.67 \times 10^{-16}$ g/cm$^3$. Very small dimensionless dissipation coefficients $\nu = \mu = 10^{-7}$ were used.

Figure 8 (top-right) shows the average over the spectra of 18 segments of the modeled energy dissipation data each with a 270 min length and 12 s cadence to match our observational data. The average spectrum has three scaling regimes with break points comparable to those of the hot 131 Å intensity spectrum and scaling exponents (0.82, 2.76, 0). While the precise values differ from those of the data, they are not far apart, and the qualitative properties are the same indicating a 1/$f$ process for the lower frequencies, strong persistence with Hurst exponent H=0.88 for the intermediate regime, and white noise at higher frequencies. The notable resemblance of the model spectra to the hot 131 Å spectra further strengthens the idea that the hot 131 Å signal can serve as a proxy for the dissipated energy input to the loop. In the calculations leading to Figure 8, the assumed width of the loops studied was nine pixels or 3.6 Mm (see Figure 1). Thus, if the loops were to undergo transverse movements of $\sim 2$ Mm or more over the 270 min observational time span, then spurious intensity fluctuations might appear. Because of its importance to our results, we have checked the robustness of the hot 131 Å mean spectrum in Figure 8 (top-left) against transverse motions of the loops by means of increasing the assumed loop width and looking for changes in the spectrum. We have recalculated the hot 131 Å spectra with assumed loop widths of 7.2 Mm and 14.4 Mm. We find that the three-scaling-regime pattern is preserved in all cases. The scaling exponents for the low-frequency 1/$f$ portion of the spectra change from 1.31 ± 0.07 for loop width 3.6 Mm to 1.32 ± 0.09 for loop width 7.2 Mm and to 1.40 ± 0.09 for loop width 14.4 Mm. These are all consistent within estimated error limits. Finally we have considered the effect of using different calibration factors when extracting the hot component of the 131 Å signal. With a coefficient $a = 0.032$ we find that the long period exponent of the average spectrum is -1.23 ± 0.08. Comparing to the case presented in Figure 8 with a = 0.044 and scaling exponent -1.31 ± 0.07 shows that these values are consistent within their error bars.

We have also investigated the intensity power spectra for the other EUV channels. No scaling ranges were found for the hot 94 Å and the 335 Å signals. Figure **8** (bottom) shows the results for the 211 Å and 171 Å signals. The scaling range in these cases and for the 193 Å signal (not shown) approximately extends over all the available frequencies: from the Nyquist frequency 2.5/min to 0.004/min. The scaling exponents for the 171 Å, 193 Å and 211 Å power spectra are $\beta$ = 2.07 ± 0.03, 2.30 ± 0.03, and 2.39 ± 0.03, respectively. These correspond to Hurst exponents 0.54, 0.65 and 0.70, respectively, indicating the presence of long-term persistence in these signals. When the assumed loop width is increased to 14.4 Mm, the scaling exponents become 1.98 ± 0.02, 2.10 ± 0.02, and 2.12 ± 0.02. So the intensity variations become more nearly Brownian in nature. The energy released into the loops in these warm channels follows a different process than the energy originally deposited (using the hot 131 Å signal as a proxy). We propose that the 94 Å fluctuations are related to the hot 131 Å input proxy by cooling processes from 10 MK to 7 MK. The 171 Å, 193 Å and 211 Å fluctuations appear to behave quite differently, and we propose that they are governed by the later evaporation of chromospheric ions into the loops where they radiate persistently. The 335 Å emission pattern appears to include both processes.

## 7. Summary of Results and Discussion

The detailed time evolution of loops A-E, for what we have identified as the inter-moss region based on the properties of the photospheric LOS magnetic field, shows a pattern of brightenings in which the hot 131 Å intensity peak was followed quickly by the hot 94 Å signal and then the cooler channels in progressive order of cooling, resembling the simulation of a high intensity nanoflare storm (Viall & Klimchuck 2011). The same result has been encountered by Petkaki et al. (2012) in the case of an isolated C1-class solar flare. In these cases we found that the temperature and electron density are functionally related in a manner compatible with an impulsive heating process and with the hot 131 Å signal preceding the temperature peak. The density increases as the loops fill with plasma which radiates in the cooler channels. When the heating impulses are closer together in time we encounter also examples of higher frequency heating with partial cascades only in the three hottest channels. Since the hot 131 Å signal has a clear association with the temperature increase we can use it as a proxy for the energy deposition. We find that the average power spectra in the hot 131 Å signals presents a $1/f^{\beta}$ form with three scaling regimes. For $f < 0.1$/min we find the scaling exponent $0.5 < \beta < 1.5$ which corresponds to $1/f$ noise. A possible mechanism to explain this type of scaling is through a model characterized by the superposition of impulses with response functions obeying a power law (Lowen & Teich 1990). This is reminiscent of the prescription for the nanoflare storm models with the difference that those use simplified triangular or square pulses. The range of scaling frequencies is compatible with the statistical analysis of UW14 who find a characteristic rate of 2-3 heating events/hr in hot 94 Å signals in active regions. Their simulations suggest that the actual rate could be much higher. However our results indicate that for $f > 0.1$/min the scaling exponent $\beta \geq 2$ indicating the presence of long term persistence. It would be very useful to calculate the power spectra of nanoflare models with different impulsive event rates.

We have found that the hybrid shell model for loop heating via the dissipation of turbulent energy presents spectra with the same scaling properties and comparable break points as the hot 131 Å data. In this context the impulsive events result from the intermittent turbulent energy deposition. We note that although this model does not describe some spatial details like the precise mechanism of magnetic reconnection, it is able to reach very large Reynolds numbers, not yet accessible in the direct numerical simulations of 3D MHD, and therefore it is very powerful in the reproduction of a very long dissipative signal (with a wide spectrum), rich in the number of intense energy releases. This has proven to be crucial for a comparison with the data. On the other hand models for 3D MHD turbulence have shown that the kind of fast magnetic reconnection that is required for flare and flare-like processes can be achieved as a consequence of the wandering of the stochastic magnetic field at sub-resolution scales (Lazarian & Vishniac 1999; Eyink, Lazarian & Vishniac 2011; Eyink et al. 2013; Browning & Lazarian 2013). The fact that the turbulent model with its scaling properties fits the observations suggests the presence of fast heating mechanisms.

**Acknowledgements**

AIA data are courtesy of NASA/SDO and the AIA science team. A.C. Cadavid acknowledges support from the Interdisciplinary Research Institute for the Sciences (IRIS) at California State University Northridge. We thank P. Testa and the anonymous referee for helpful suggestions.